\documentclass{PoS}

\usepackage{epsf}
\usepackage{amssymb}
\usepackage{amsmath}
\usepackage{amsfonts}
\usepackage{cite}
\usepackage[small]{caption2}

\usepackage[T1]{fontenc}
\usepackage{psfrag,epsfig,graphicx,graphics}

\newcommand{\be}{\begin{equation}}
\newcommand{\beq}{\begin{equation}}
\newcommand{\ee}{\end{equation}}
\newcommand{\eq}{\end{equation}}
\newcommand{\eeq}{\end{equation}}

\newcommand{\bea}{\begin{eqnarray}}
\newcommand{\eea}{\end{eqnarray}}

\def\slashchar#1{\setbox0=\hbox{$#1$}
   \dimen0=\wd0
   \setbox1=\hbox{/} \dimen1=\wd1
   \ifdim\dimen0>\dimen1
      \rlap{\hbox to \dimen0{\hfil/\hfil}}
      #1
   \else
      \rlap{\hbox to \dimen1{\hfil$#1$\hfil}}
      /gdatdafinal2.tex
   \fi}

\title{On chiral-odd Generalized Parton Distributions  
}

\ShortTitle{On chiral-odd Generalized Parton Distributions  
}

\author{M. El Beiyad\\
CPHT, {\'E}cole Polytechnique, CNRS, 91128 Palaiseau Cedex, France \ {\em \&} \\
LPT, Universit{\'e} Paris-Sud, CNRS, 91405 Orsay, France
\\
E-mail: \email{Mounir.Elbeiyad@cpht.polytechnique.fr}}

\author{B.~Pire\\
CPHT, {\'E}cole Polytechnique, CNRS, 91128 Palaiseau Cedex, France\\
E-mail: \email{pire@cpht.polytechnique.fr}}

\author{M.~Segond\\
Institut f\"ur Theoretische
Physik, Universit\"at Leipzig,  D-04009 Leipzig, Germany\\
E-mail: \email{Mathieu.Segond@itp.uni-leipzig.de}}

\author{L.~Szymanowski\\
Soltan Institute for Nuclear Studies, PL-00-681 Warsaw, Poland\\
E-mail: \email{Lech.Szymanowski@fuw.edu.pl}}

\author{\speaker{S.~Wallon}%\thanks{A footnote may follow.}
\\
LPT, Universit{\'e} Paris-Sud, CNRS, 91405 Orsay, France \ {\em \&} \\
UPMC Univ. Paris 06, facult\'e de physique, 4 place Jussieu, 75252 Paris Cedex  05, France\\
        E-mail: \email{wallon@th.u-psud.fr}}

\abstract{
While the chiral-even content of the nucleon is better and better studied, the chiral-odd transversity generalized parton distributions (GPDs) remain almost unknown. These GPDs can be accessed experimentally through the exclusive photoproduction process $\gamma + N \to \pi + \rho + N',$ in the kinematics where the meson pair has a large invariant mass and the final nucleon has a small transverse momentum, provided the vector meson is produced in a transversally polarized state. We calculate perturbatively the scattering amplitude at leading order in $\alpha_s.$ We build a simple model for the dominant transversity GPD $H_T(x,\xi, t)$ based on the concept of double distribution. 
We estimate  the unpolarized differential cross section for this process  in the kinematics of
the Jlab and COMPASS experiments. Counting rates show that the experiment
looks feasible with the real and quasi real photon beam characteristics expected at 
JLab@12 GeV, and with the quasi real photon beam in the COMPASS experiment.
}

\FullConference{35th International Conference of High Energy Physics\\
                 July 22-28, 2010\\
                 Paris, France}

\begin{document}

\section{Chiral-odd GPDs and factorization}
\label{Sec:Int}

\begin{figure}[h]
\begin{center}
\psfrag{z}{\begin{small} $\hspace{-.1cm}z$ \end{small}}
\psfrag{zb}{\raisebox{-0cm}{ \begin{small}$\hspace{-.1cm}\bar{z}$\end{small}} }
\psfrag{gamma}{\raisebox{+.1cm}{\hspace{-0.2cm} $\,\gamma$} }
\psfrag{pi}{$\!\!\pi$}
\psfrag{rho}{\raisebox{-.05cm}{$\,\rho_T$}}
\psfrag{TH}{\hspace{-0.2cm} $T_H$}
\psfrag{tp}{\raisebox{.6cm}{$t'$}}
\psfrag{s}{\hspace{0.4cm} $s$}
\psfrag{Phi}{\raisebox{-.05cm}{ \hspace{-0.25cm} $\phi$}}
\hspace{-0.7cm}
\vspace{0.25cm}
\raisebox{.1cm}{\includegraphics[width=5.3cm]{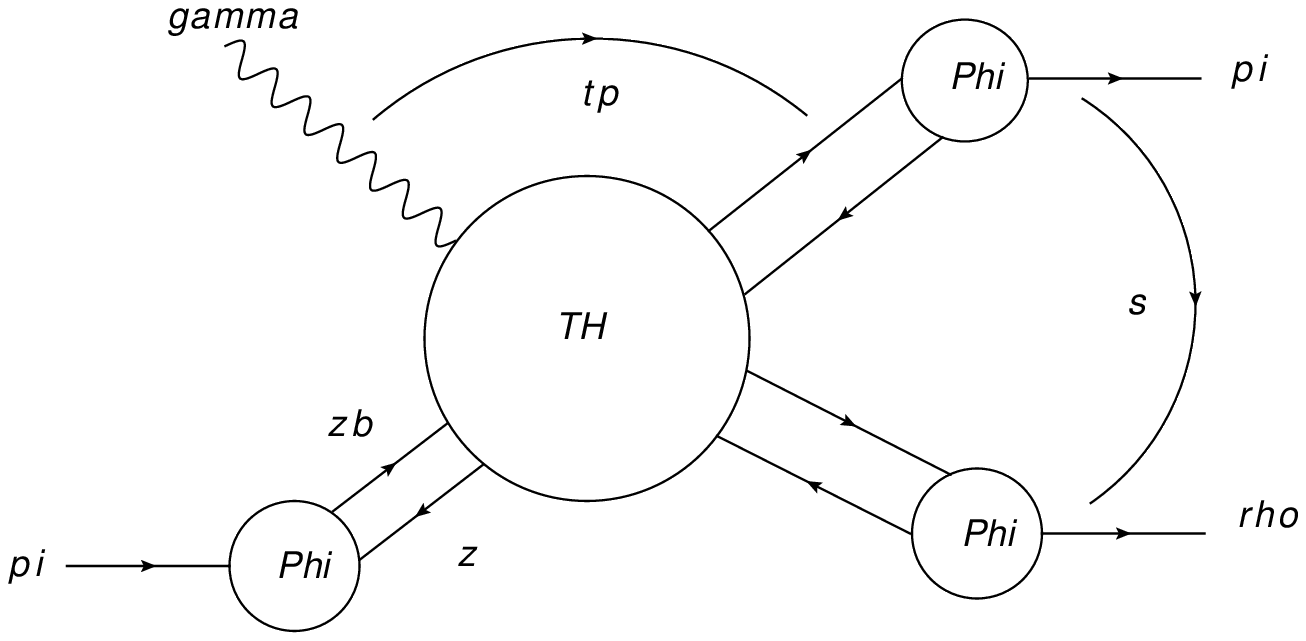}}~~\hspace{1.2cm}
\psfrag{piplus}{$\,\pi$}
\psfrag{rhoT}{\raisebox{-.05cm}{$\,\rho_T$}}
\psfrag{M}{\hspace{-0.3cm} \begin{small} $M^2_{\pi \rho}$ \end{small}}
\psfrag{x1}{\hspace{-0.7cm} \begin{small}  $x+\xi $  \end{small}}
\psfrag{x2}{ \hspace{-0.2cm}\begin{small}  $x-\xi $ \end{small}}
\psfrag{N}{ \hspace{-0.4cm} $N$}
\psfrag{GPD}{ \hspace{-0.6cm}  $GPDs$}
\psfrag{Np}{$N'$}
\psfrag{t}{ \raisebox{-.1cm}{ \hspace{-0.5cm} $t$ }}
\psfrag{tp}{\raisebox{.4cm}{{\begin{small}     $t'$       \end{small}}}}
 \raisebox{-0.4cm}{\includegraphics[width=4.9cm]{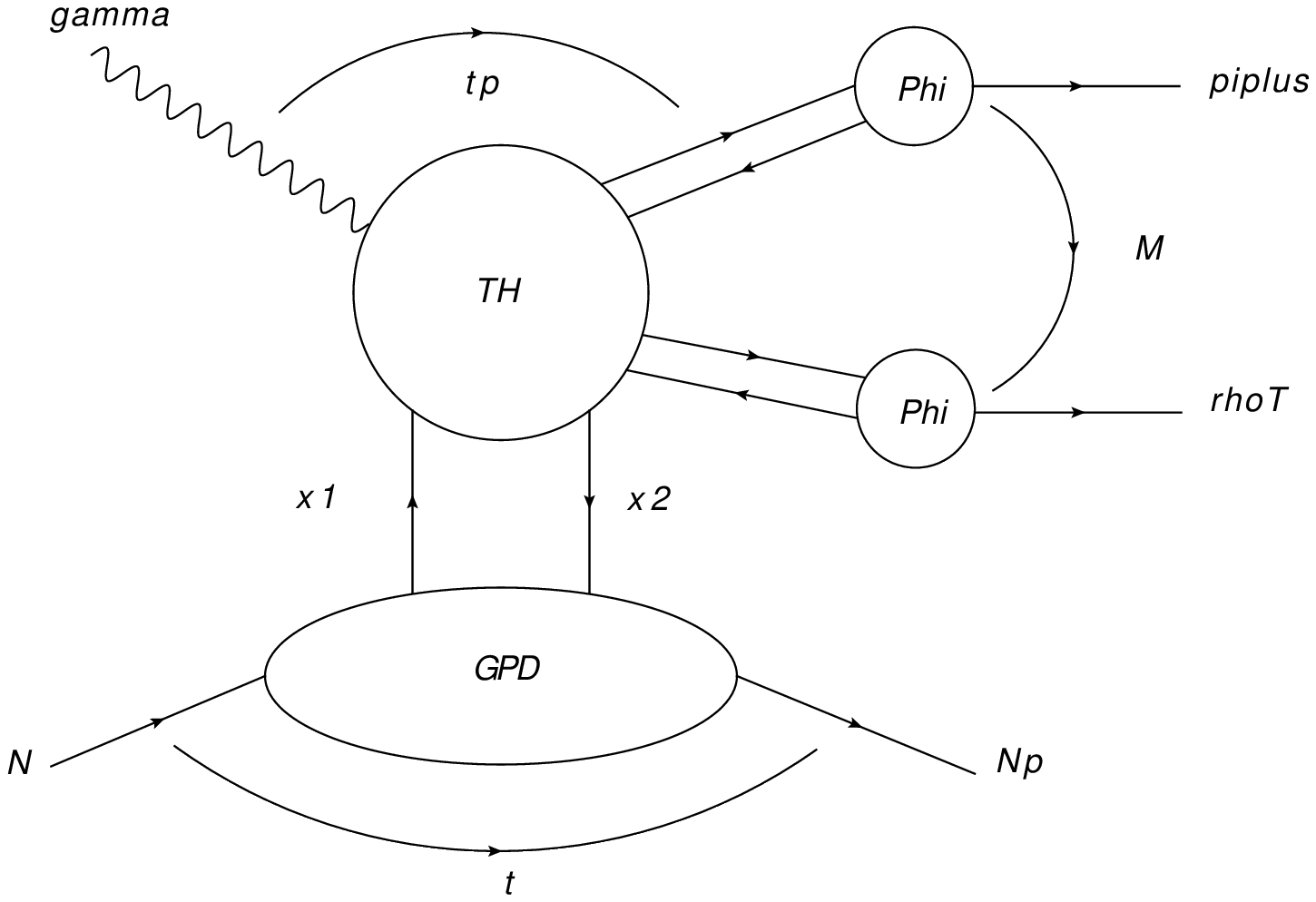}}
\end{center}
\vspace{-.7cm}
\caption{a) (left) Factorization of the amplitude for $\gamma + \pi \rightarrow \pi + \rho $ at large $s$ and fixed angle; %(i.e. fixed ratio $t'/s$)
 b) (right) replacing one DA by a GPD leads to the factorization of the amplitude  for $\gamma + N \rightarrow \pi + \rho +N'$ at large $M_{\pi\rho}^2$.}
\label{feyndiag}
%\end{center}
\end{figure}

Among the leading twist hadronic observables, 
transversity quark distributions in the nucleon remain  the most unknown ones. Indeed their chiral-odd character enforces their decoupling in most hard amplitudes at the leading twist level. After the pioneering studies of Ref.~\cite{tra}, much work \cite{Barone} has  been devoted to the study of many channels but experimental difficulties have challenged the most promising ones.
Besides, tremendous progress has been recently witnessed on the \nolinebreak QCD  description of hard exclusive processes, in terms of generalized parton distributions (GPDs) describing the 3-dimensional content of hadrons. Access to the chiral-odd transversity GPDs~\cite{defDiehl},
%\nolinebreak 
noted   $H_T$, $E_T$, $\tilde{H}_T$, $\tilde{E}_T$, has however turned out to be even more challenging~\cite{DGP} than the usual transversity distributions: one photon or one meson electroproduction leading twist amplitudes are insensitive  to transversity GPDs.  
As initiated in Ref.~\cite{IPST}, we here study the leading twist  contribution  to processes where more mesons are present in the final state\footnote{A similar strategy has also been advocated recently in Ref.~\cite{kumano} for chiral-even GPDs.
%to enlarge the number of processes which could be used to extract information on %chiral-even GPDs
}; the hard scale which allows to probe the short distance structure of the nucleon
is  $s=M_{\pi \rho}^2\, \sim |t'|$ in the fixed angle regime.
%
% is now the invariant mass of the meson pair, provided it is of the same order as related to the large transverse momentum transmitted to  each final meson.
%
The process \nolinebreak of Ref.~\cite{IPST}  was the high energy photo (or electro) diffractive production of two vector mesons, the  hard probe being the virtual "Pomeron" exchange (the hard scale was the virtuality of \nolinebreak this pomeron), in analogy with the virtual photon exchange occuring in the deep electroproduction of a meson. 
We study here \cite{PLB} a process involving %a 3-body final state with 
a transversely polarized $\rho$ meson in a 3-body final  state:
\begin{equation}
\gamma + N \rightarrow \pi + \rho_T + N'\,.
\label{process}
\end{equation}
It is a priori sensitive to chiral-odd GPDs due to the chiral-odd character of the leading twist distribution amplitude (DA) of $\rho_T$.  The estimated rate depends of course much on the magnitude of the chiral-odd GPDs. Not much is known about them, but model calculations have been developed in Refs.~\cite{IPST,Sco,Pasq,othermodels} and  a few moments have been computed on the lattice~\cite{lattice}.
To factorize the amplitude of this process we use  the now classical proof of the factorization of exclusive scattering at fixed angle and large energy~\cite{LB}. The amplitude for the process $\gamma + \pi \rightarrow \pi + \rho $ is written as the convolution of mesonic DAs  and a hard scattering subprocess amplitude $\gamma +( q + \bar q) \rightarrow (q + \bar q) + (q + \bar q) $ with the meson states replaced by collinear quark-antiquark pairs. This is described in Fig.\ref{feyndiag}a. The absence of any pinch singularities (which is the weak point of the proof for the generic case $A+B\to C+D$) has been proven in Ref.~\cite{FSZ} for the case  of interest here.
We then extract from the factorization procedure of the deeply virtual Compton scattering amplitude near the forward region the right to replace in Fig.\ref{feyndiag}a the lower left meson DA by a $N \to N'$ GPD, and thus get Fig.\ref{feyndiag}b. 
The needed skewness parameter $\xi$  is written in terms of the meson pair squared invariant mass
$M^2_{\pi\rho}$ as
\begin{equation}
\label{skewedness}
\xi = \frac{\tau}{2-\tau} ~~~~,~~~~\tau =
\frac{M^2_{\pi\rho}}{S_{\gamma N}-M^2}\,.
\end{equation}
Indeed the same collinear factorization property bases the validity of the leading twist approximation which either replaces the meson wave function by its DA or the $N \to N'$ transition by nucleon GPDs. A slight difference is that light cone fractions ($z, 1- z$) leaving the DA are positive, while the corresponding fractions ($x+\xi,\xi-x$) may be positive or negative in the case of the GPD. A  
Born order calculation  show that this difference does not ruin the factorization property.
In order \nopagebreak
for the factorization of a partonic amplitude to be valid, and the
leading twist calculation to be suf-
\pagebreak

\noindent
\vspace{0.1cm}
%
%%%%%%%%%%%%%%%%%%     FIGURE 1          %%%%%%%%%%%%%%%%%%%%%%%%%%%%
%%%%%%%%%%%%%%%%%%%%%%%%%%%%%%%%%%%%%%%%%%%%%%%%%%%%%%%
%
\noindent
ficient, one should avoid the dangerous
kinematical regions where a small momentum transfer is exchanged in the
upper blob, namely small $t' =(p_\pi -p_\gamma)^2$ or small
$u'=(p_\rho-p_\gamma)^2$, and the resonance regions for each  of the
invariant squared masses $(p_\pi +p_{N'})^2 , (p_\rho +p_{N'})^2, (p_\pi +p_\rho)^2\,.$

\section{The scattering amplitude}
\label{Sec:scattering}

%%%%%%%%%%%%%%%%%%     FIGURE 2          %%%%%%%%%%%%%%%%%%%%%%%%%%%%
\begin{figure}[!h]
\centerline{$\begin{array}{cc}
\psfrag{fpi}{$\hspace{-.015cm}\phi_\pi$}
\psfrag{fro}{$\hspace{-.015cm}\phi_\rho$}
\psfrag{z}{\begin{small} $z$ \end{small}}
\psfrag{zb}{\raisebox{-.2cm}{ \begin{small}$\hspace{-.3cm}-\bar{z}$\end{small}} }
\psfrag{v}{\begin{small} $v$ \end{small}}
\psfrag{vb}{\raisebox{-.1cm}{ \begin{small}$\hspace{-.4cm}-\bar{v}$\end{small}} }
\psfrag{gamma}{$\,\gamma$}
%\psfrag{pi}{$\,\pi^+$}
\psfrag{pi}{$\,\pi$}
%\psfrag{rho}{$\,\rho^0_T$}
\psfrag{rho}{$\,\rho_T$}
\psfrag{N}{$N$}
\psfrag{Np}{$\,N'$}
\psfrag{H}{\raisebox{-.1cm}{\hspace{-0.3cm} $H_T(x,\xi,t)$}}
\psfrag{p1}{\!\!\begin{small}     $p_1$       \end{small}}
\psfrag{p2}{\!\begin{small} $p_2$ \end{small}}
\psfrag{p1p}{\hspace{-1.2cm}  \begin{small}  $p_1'=(x+\xi) p$  \end{small}}
\psfrag{p2p}{\hspace{-0.2cm} \begin{small}  $p_2'=(x-\xi) p$ \end{small}}
\psfrag{q}{\begin{small}     $\hspace{-.5cm} q$       \end{small}}
\psfrag{ppi}{\begin{small} $p_\pi$\end{small}}
\psfrag{prho}{\begin{small} $p_\rho$\end{small}}
\includegraphics[width=5.1cm]{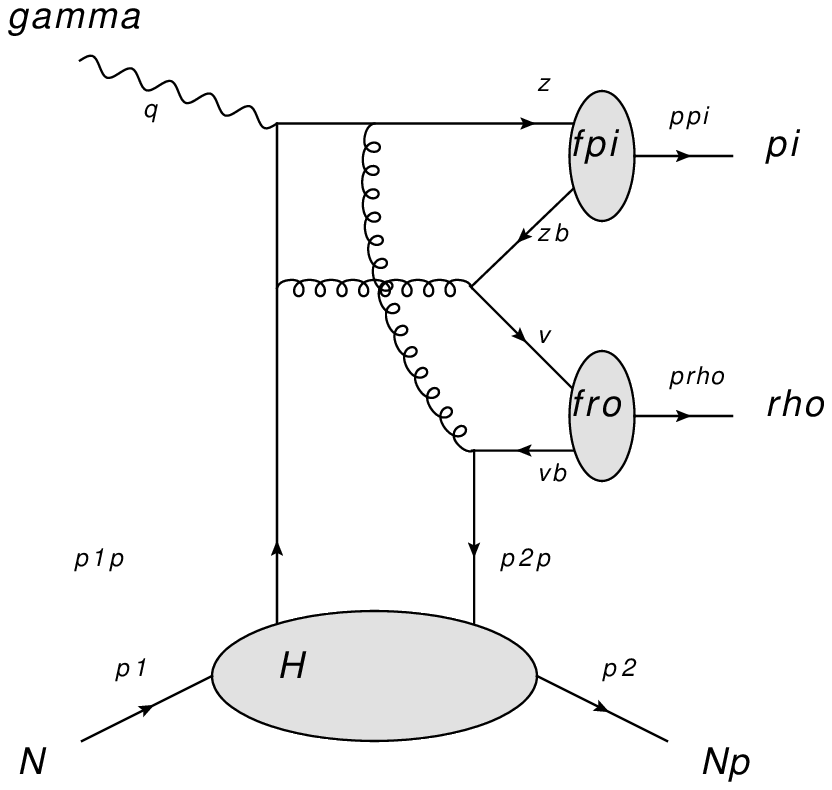}&\hspace{1cm}
\psfrag{fpi}{$\hspace{-.015cm}\phi_\pi$}
\psfrag{fro}{$\hspace{-.015cm}\phi_\rho$}
\psfrag{z}{\begin{small} $z$ \end{small}}
\psfrag{zb}{\raisebox{-.2cm}{ \begin{small}$\hspace{-.3cm}-\bar{z}$\end{small}} }
\psfrag{v}{\begin{small} $v$ \end{small}}
\psfrag{vb}{\raisebox{-.1cm}{ \begin{small}$\hspace{-.4cm}-\bar{v}$\end{small}} }
\psfrag{gamma}{$\,\gamma$}
%\psfrag{pi}{$\,\pi^+$}
\psfrag{pi}{$\,\pi$}
%\psfrag{rho}{$\,\rho^0_T$}
\psfrag{rho}{$\,\rho_T$}
\psfrag{N}{$N$}
\psfrag{Np}{$\,N'$}
\psfrag{H}{\raisebox{-.1cm}{\hspace{-0.3cm} $H_T(x,\xi,t)$}}
\psfrag{p1}{\!\begin{small}     $p_1$       \end{small}}
\psfrag{p2}{\begin{small} $p_2$ \end{small}}
\psfrag{p1p}{\hspace{-1.2cm}  \begin{small}  $p_1'=(x+\xi) p$  \end{small}}
\psfrag{p2p}{\hspace{-0.2cm} \begin{small}  $p_2'=(x-\xi) p$ \end{small}}
\psfrag{q}{\begin{small}     $\hspace{-.5cm}q$       \end{small}}
\psfrag{ppi}{\begin{small} $p_\pi$\end{small}}
\psfrag{prho}{\begin{small} $p_\rho$\end{small}}
\includegraphics[width=5.1cm]{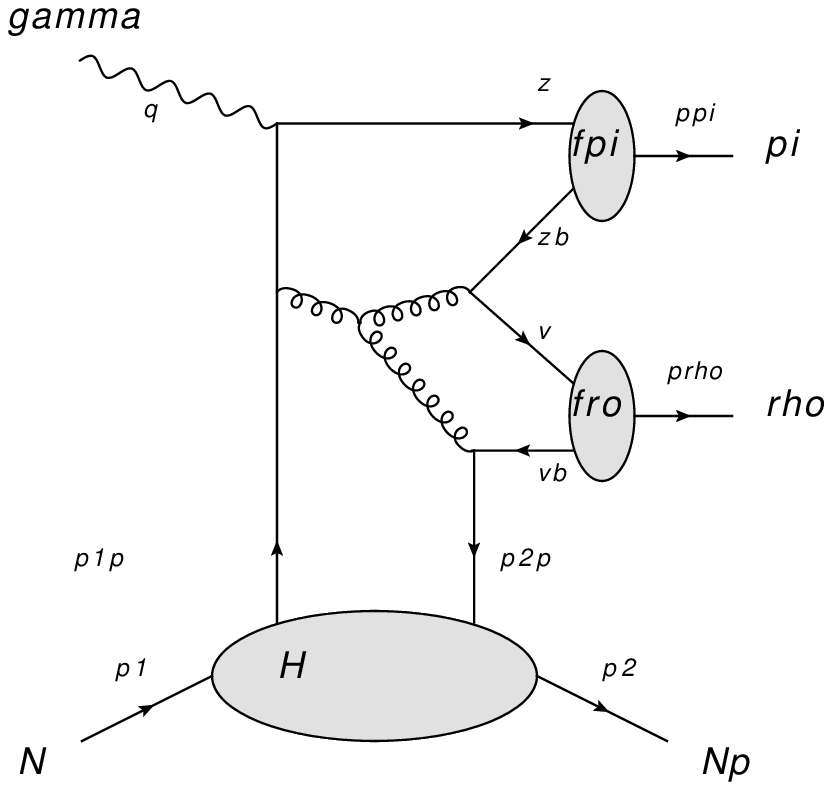}
\end{array}$}
\caption{Two representative diagrams without (left) and with (right) three gluon coupling.}
\label{feyndiageued}
\end{figure}
%%%%%%%%%%%%%%%%%%%%%%%%%%%%%%%%%%%%%%%%%%%%%%%%%%%%%%%
The scattering amplitude of the process (\ref{process}) is written in the factorized form :
\begin{equation}
\label{AmplitudeFactorized}
\mathcal{A}(t,M^2_{\pi\rho},p_T)  = \int_{-1}^1dx\int_0^1dv\int_0^1dz\ T^q(x,v,z) \, H^{q}_T(x,\xi,t)\Phi_\pi(z)\Phi_\bot(v)\,,
\end{equation}
where
$T^q$ is the hard part of the amplitude and
the transversity GPD of a parton $q$  in the nucleon target which dominates at small momentum transfer is defined by~\cite{defDiehl}
\[
%\begin{equation}
\langle N'(p_2),\lambda'|\bar{q}\left(-\frac{y}{2}\right)\sigma^{+j}\gamma^5 q \left(\frac{y}{2}\right)|N(p_1),\lambda \rangle  = \bar{u}(p',\lambda')\sigma^{+j}\gamma^5u(p,\lambda)\int_{-1}^1dx\ e^{-\frac{i}{2}x(p_1^++p_2^+)y^-}H_T^q\,,
%\label{defGPD}
%\end{equation}
\]
where $\lambda$ and $\lambda'$ are the light-cone helicities of the nucleon $N$ and $N'$.
The chiral-odd  DA for the  transversely polarized meson vector $\rho_T$,  is defined, in leading twist 2, by the matrix element %~\cite{defrho}
\[
%\begin{equation}Numerical e
\langle 0|\bar{u}(0)\sigma^{\mu\nu}u(x)|\rho^0_T(p,\epsilon_\pm) \rangle =\frac{i}{\sqrt{2}} (\epsilon^\mu_{\pm}(p) p^\nu - \epsilon^\nu_{\pm}(p) p^\mu)f_\rho^\bot\int_0^1du\ e^{-iup\cdot x}\ \phi_\bot(u)\,,
%\label{defDArho}
%\end{equation}
\]
where $\epsilon^\mu_{\pm}(p_\rho)$ is the $\rho$-meson transverse polarization and with $f_\rho^\bot$ = 160 MeV. 
This process is described by 62 Feynman diagrams (see Fig.\ref{feyndiageued}). The scattering amplitude gets both a real and an 
imaginary parts. Integrations over $v$ and $z$ have been  
done analytically whereas numerical methods are used for the integration over $x$.\pagebreak

\noindent

\section{Results}
\label{sec:results}

\begin{figure}[!h]
\begin{center}
\includegraphics[width=7cm]{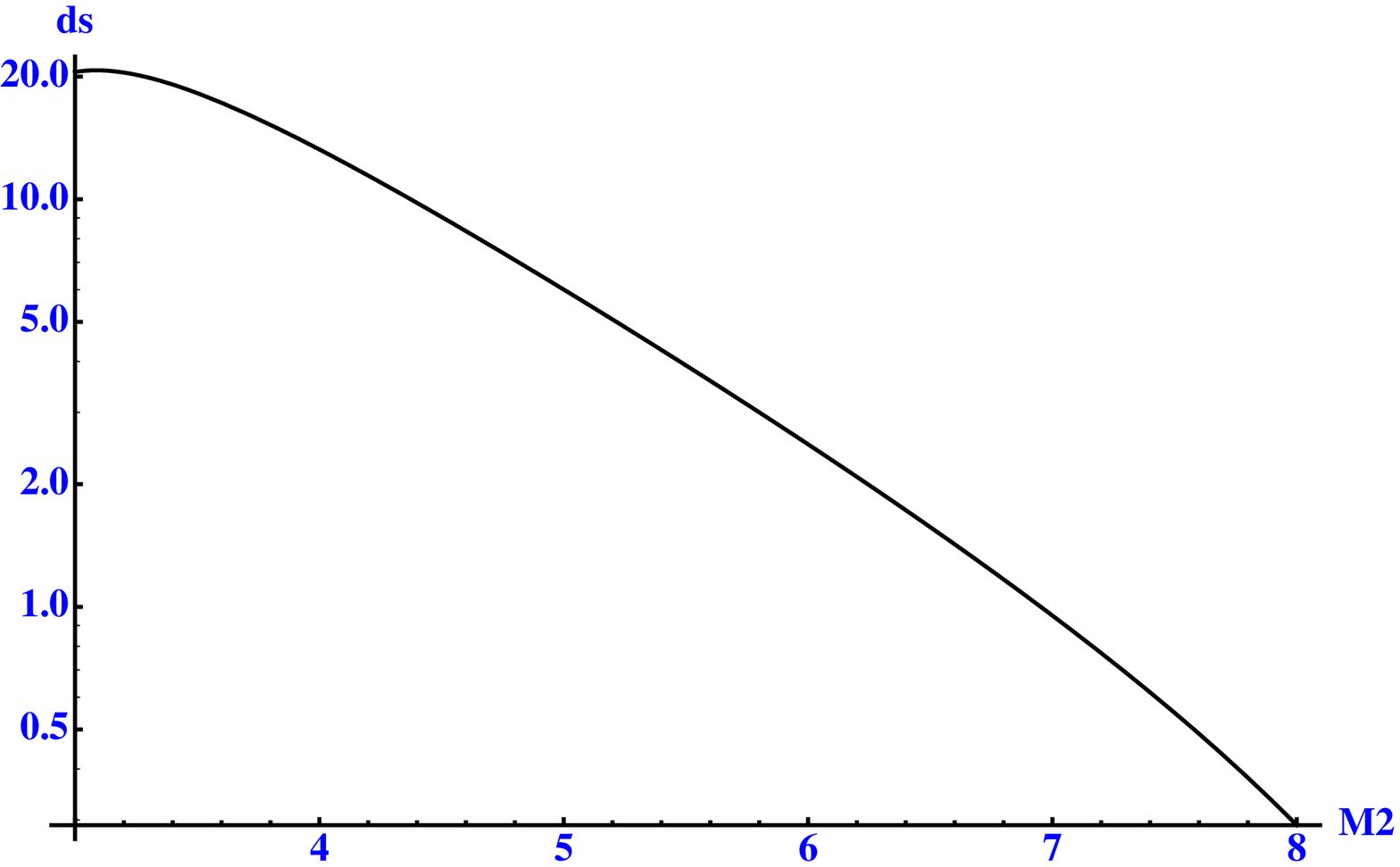} \quad \includegraphics[width=7cm]{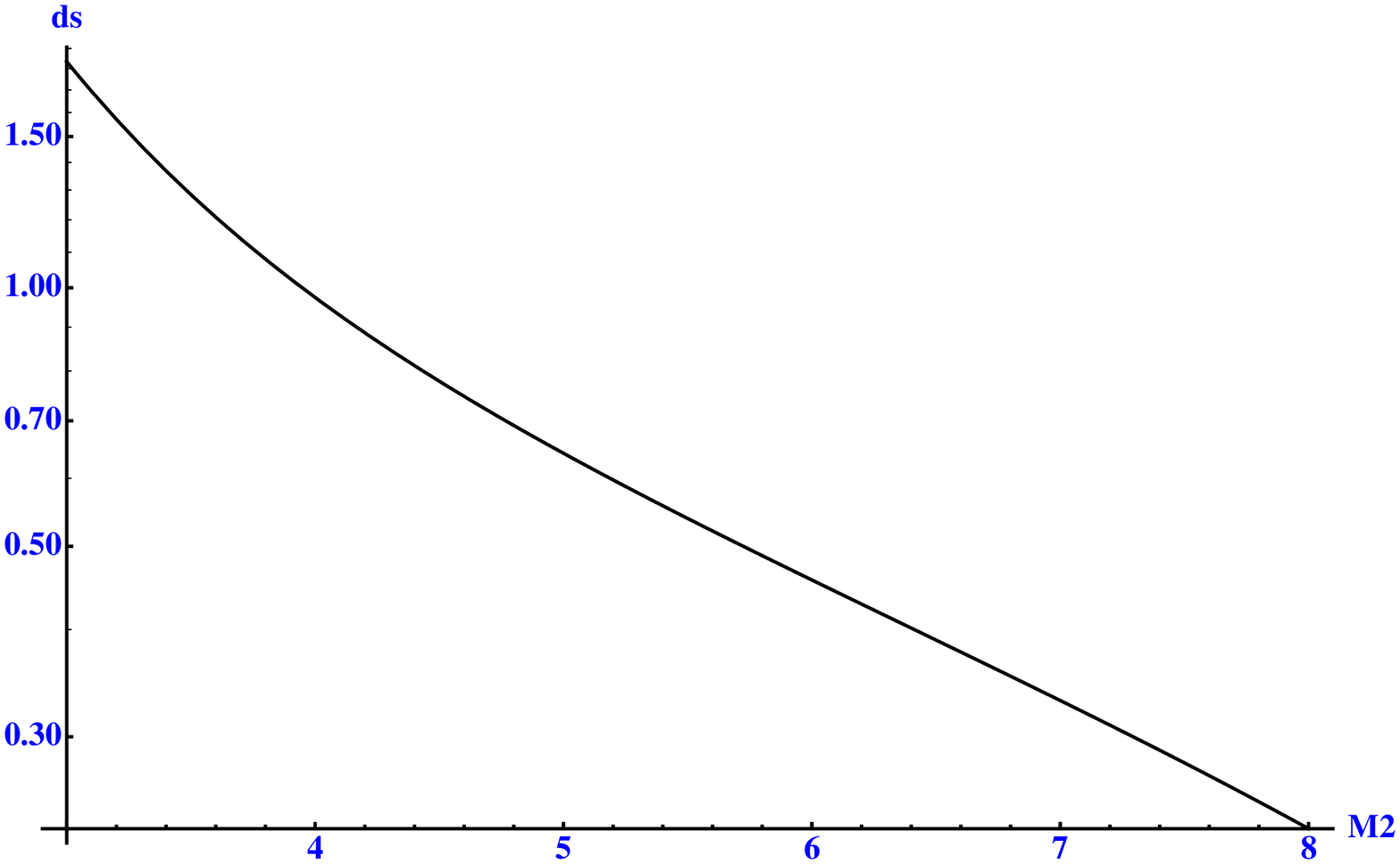} 
\vspace{.3cm}
\caption{$\displaystyle \frac{d \sigma}{d M_{\pi \rho}^2}$%(\ref{difcrosec2})
(nb.GeV$^{-2}$) %in dependence of $M^2_{\pi\rho}$ 
at $S_{\gamma N}$ = 20 GeV$^2$ (left) and $S_{\gamma N}$ = 100 GeV$^2$ (right).}
\label{result20and100}
\end{center}
\end{figure}

Various observables can be calculated with this amplitude. We stress that even the unpolarized differential cross-section $\frac{d\sigma}{dt \,du' \, dM^2_{\pi\rho}}$
is sensitive to the transversity GPD.
To estimate the rates, we modelize the dominant transversity GPD $H_T^q(x,\xi,t)$ ($q=u,\ d$) in
terms of double distributions 
\begin{equation}
\label{DDdef}
H_T^q(x,\xi,t=0) = \int_\Omega d\beta\, d\alpha\ \delta(\beta+\xi\alpha-x)f_T^q(\beta,\alpha,t=0)
\,,
\end{equation}
where $f_T^q$ is the quark transversity double distribution written as 
\begin{equation}
\label{DD}
f_T^q(\beta,\alpha,t=0) = \Pi(\beta,\alpha)\,\delta \, q(\beta)\Theta(\beta) -
\Pi(-\beta,\alpha)\,\delta \bar{q}(-\beta)\,\Theta(-\beta)\,,
\end{equation}
where $ \Pi(\beta,\alpha) = \frac{3}{4}\frac{(1-\beta)^2-\alpha^2}{(1-\beta)^3}$ is a profile
function and $\delta q$, $\delta \bar{q}$ are the quark and antiquark transversity parton
distribution functions  of Ref.~\cite{Anselmino}. The $t$-dependence
of these chiral-odd GPDs - and its Fourier transform in terms of the transverse
localization of quarks in the proton \cite{impact} - is very interesting but completely unknown.
We
 describe it in a simplistic way using a dipole form factor: 
\begin{equation}
\label{t-dep}
H^q_T(x,\xi,t) = H^q_T(x,\xi,t=0)\times \frac{C^2}{(t  - C)^2} \qquad (C=.71~{\rm GeV}^2)\,.
\end{equation}
% with  $C=.71~$GeV$^2$.  
In Fig.~\ref{result20and100},  we show the $M^2_{\pi\rho}$ dependence of the
differential cross section $\displaystyle d \sigma/d M_{\pi \rho}^2.$%(\ref{difcrosec2}).
%
%%%%%%%%%%%%%%%%%%     FIGURE 10         %%%%%%%%%%%%%%%%%%%%%%%%%%%%
\vspace{.4cm}
\psfrag{ds}{\raisebox{.4cm}{{\hspace{-.6cm} $ \frac{d \sigma}{d M_{\pi \rho}^2}$
\hspace{0cm}{\small (nb.GeV$^{-2}$)}}}}
\psfrag{M2}{\small\raisebox{-.7cm}{{\hspace{-2.5cm}$M_{\pi \rho}^2$(GeV$^2$)}}}
%%%%%%%%%%%%%%%%%%%%%%%%%%%%%%%%%%%%%%%%%%%%%%%%%%%%%%%
%
% %%%%%%%%%%%%%%%%%%     FIGURE 10         %%%%%%%%%%%%%%%%%%%%%%%%%%%%
% \psfrag{ds}{\raisebox{.5cm}{{\hspace{-.6cm}$\displaystyle \frac{d \sigma}{d M_{\pi \rho}^2}$
% \hspace{0cm}{ (nb.GeV$^{-2}$)}}}}
% \psfrag{M2}{\raisebox{-.7cm}{{\hspace{-2.5cm}$M_{\pi \rho}^2$(GeV$^2$)}}}
% \begin{figure}[!h]
% \begin{center}
% \includegraphics[width=12cm]{dsigma_M2_JLab_D_fit_new.eps}
% \vspace{.3cm}
% \caption{$M^2_{\pi\rho}$ dependence of the differential cross section $\displaystyle \frac{d \sigma}{d M_{\pi \rho}^2}$%(\ref{difcrosec2})
% (nb.GeV$^{-2}$) at $S_{\gamma N}$ = 20 GeV$^2$.}
% \label{result20}
% \end{center}
% \end{figure}
% %%%%%%%%%%%%%%%%%%%%%%%%%%%%%%%%%%%%%%%%%%%%%%%%%%%%%%%
% %
% %%%%%%%%%%%%%%%%%%     FIGURE 11        %%%%%%%%%%%%%%%%%%%%%%%%%%%%
% \begin{figure}[!h]
% \begin{center}
% \vspace{.5cm}
% \includegraphics[width=12cm]{dsigma_M2_Cp_bE_fit_new.eps}
% \vspace{.3cm}
% \caption{$M^2_{\pi\rho}$ dependence of the differential cross section (\ref{difcrosec2})
% (nb.GeV$^{-2}$) at $S_{\gamma N}$ = 100 GeV$^2$.}
% \label{result100}
% \end{center}
% \end{figure}
% %%%%%%%%%%%%%%%%%%%%%%%%%%%%%%%%%%%%%%%%%%%%%%%%%%%%%%%
Let us first discuss the specific case of muoproduction with the COMPASS experiment at CERN.
Integrating differential cross sections on $t$, $u'$ and $M^2_{\pi\rho}$, with justified cuts on
$t'$ and $u'$ and $M^2_{\pi\rho} > 3$ GeV$^2$,  leads to an estimate of the cross sections for the
photoproduction of a $\pi^+\rho^0_T$ pair at high energies such as
\begin{equation}
\label{crossec1}
\sigma_{\gamma N\to \pi^+\rho^0_TN'}(S_{\gamma N} = 100\ GeV^2) \simeq 3\ \textrm{nb} .
\end{equation}
The virtuality $Q^2$ of the exchanged photon plays no crucial role in our process, and the virtual photoproduction cross section is almost $Q^2$-independent if we choose to select events in a
sufficiently
 narrow $Q^2-$window ($.02<Q^2<1 $ GeV$^2$), which is legitimate since the
effective photon flux is
strongly peaked at very low values of $Q^2$.  This yields a rate
sufficient to get an estimate of the transversity GPDs in the region of small $\xi$ ($\sim
0.01$). In the lower energy domain, which will be 
studied in details at JLab, and with the same cuts
as above, estimates of the cross sections  are:
\pagebreak
 
\noindent 
\begin{equation}
\label{crossec}
\sigma_{\gamma N\to \pi^+\rho^0_TN'}(S_{\gamma N} = 10\ GeV^2) \simeq 15\ \textrm{nb} \qquad
\sigma_{\gamma N\to \pi^+\rho^0_TN'}(S_{\gamma N} = 20\ GeV^2) \simeq 33\ \textrm{nb}.
\end{equation}
% At JLab@12GeV, a detailed analysis should thus be
% possible due to the high electron beam luminosity.
%
The high electron beam luminosity of JLab@12GeV should thus allow for
a detailed analysis.

In conclusion,  the process discussed here is promissing
in order to get access to the
 chiral-odd transversity GPDs, both at 
low real or almost real photon energies within the JLab@12GeV upgraded facility,   both in Hall B and Hall D, and at higher photon energies with the  COMPASS experiment. 
We note that our proposal is 
 built on known
leading twist factorization theorems, contrarily to other attempts to access transversity GPDs
\cite{liuti}. At large center of mass energy, for chiral-even $t-$channel exchange, hard exclusive vector meson production beyond leading twist has been consistently described in Ref.~\cite{us}. This might be extended for chiral-odd quantities.
%A full treatment of hard exclusive meson production beyond leading twist is still an open problem, except at large energy \cite{us}.

%for which
%a $k_T$-factorization based description is now available \cite{us}.

%This work is partly 
Work supported
%by the French-Polish scientific agreement Polonium 7294/R08/R09,  the ECO-NET program, contract 18853PJ,
by the grants ANR-06-JCJC-0084, Polish  N202 249235,  DFG (KI$-623/4$).

%For figures, we recommend using the wrapfig package to save space. 

%To insert a figure surrounded by text (with the help of the wrapfig and

%graphicx packages---please, do avoid using either the epsfig or epsf packages as they are now considered obsolete).

\end{document}